\documentstyle[aps,prb,epsf,floats]{revtex}
\begin{document}
\twocolumn[\hsize\textwidth\columnwidth\hsize\csname@twocolumnfalse%
\endcsname
\title{Low temperature spin diffusion in the \\
one-dimensional quantum $O(3)$ nonlinear $\sigma$-model}
\author{Subir Sachdev and Kedar Damle}
\address{Department of Physics, Yale University\\
P.O. Box 208120,
New Haven, CT 06520-8120}
\date{October 11, 1996}
\maketitle

\begin{abstract}
An effective, low temperature, classical model for spin transport in the
one-dimensional, gapped, quantum $O(3)$ non-linear $\sigma$-model is
developed. Its correlators are obtained by a mapping to a model solved
earlier by Jepsen. We obtain universal functions for the
ballistic-to-diffusive crossover and the value of the spin diffusion
constant, and these are claimed to be exact at low temperatures.
Implications for experiments on one-dimensional insulators with a spin gap
are noted.
\end{abstract}
]

Over the past decade, a large number of one-dimensional, insulating, Heisenberg antiferromagnets
with a zero temperature ($T$)  spin gap have been studied: these include integer spin
chains~\cite{taki,shimizu} and half-integer spin-ladder systems~\cite{azuma}.
In the large spin $S$ limit, the low energy properties
of these compounds are  described~\cite{haldane} by the one-dimensional quantum $O(3)$
non-linear
$\sigma$-model (without any topological term), and there is evidence~\cite{soren}
that the mapping to this continuum model is quantitatively accurate even for the
 $S=1$ spin chain. Theoretically, much is
known about the quantum field theory of the
$\sigma$-model~\cite{polyakov,zama,smirnov}, and this information has been
valuable in understanding the properties of the spin chains. The low energy
spectrum of the $\sigma$-model consists of a triplet of massive particles,
and their ballistic propagation describes many exactly known dynamic
correlations at $T=0$. For $T>0$ however, exact results have so far been
limited to static, thermodynamic observables~\cite{tsvelik}.

In this paper, we obtain dynamic, non-zero $T$ correlators using a
semiclassical method~\cite{apy}: we claim that all of our results are
asymptotically exact at low $T$, but this has not been rigorously
established. We present universal functions which describe the crossover
from ballistic spin transport at short scales, to {\em diffusive\/}
behavior at the longest scales; as a bi-product, these functions yield the
exact value of the spin diffusion constant. The nature of spin transport
for any small $T>0$ is therefore qualitatively different from that at
$T=0$.

The imaginary time ($\tau$) action of the $\sigma$-model is
\begin{displaymath}
{\cal A} = \frac{c}{2g} \int_0^{1/T} d\tau dx \left[ (\partial_x n_{\alpha}
)^2 + \frac{1}{c^2}(\partial_{\tau} n_{\alpha} - i
\epsilon_{\alpha \beta \gamma} H_{\beta} n_{\gamma}
)^2  \right]
\end{displaymath}
where $x$ is the spatial co-ordinate, $\alpha,\beta,\gamma = 1,2,3$ are $O(3)$ vector
indices over which there is an implied summation, $\epsilon_{\alpha\beta\gamma}$ is the
totally antisymmetric tensor, $c$ is a velocity,
$H_{\alpha}$ is an external magnetic field, and the partition function is obtained
by integrating over the unit vector vield $n_{\alpha} (x,\tau)$, with
$n_{\alpha}^2 (x,\tau) = 1$.
We use units in which $\hbar=k_B=1$ and have absorbed a factor of the
electronic magnetic moment, $g_e \mu_B$, into the definition of the field $H$.
The dimensionless coupling constant $g$ is determined
by the underlying lattice antiferromagnet
at the momentum scale $\Lambda \sim \mbox{inverse lattice spacing}$ to
be $g \sim 1/S$. We shall only be interested in the physics at
length scales $\gg \Lambda^{-1}$ and time scales $\gg (c \Lambda)^{-1}$;
this physics is {\em universally\/} characterized by the dimensionful parameters $c$, $H$, $T$,
and $\Delta$, the energy gap at $T=H=0$. The magnitude of
$\Delta$ is determined by non-universal lattice scale physics
($\Delta \sim c \Lambda e^{-2\pi/g}$ for small $g$). However, the long distance physics
depends on these lattice scale effects only through the value of $\Delta$,
and has no direct dependence on $g$ or $\Lambda$.

We shall study correlators of the magnetization density, $M_{\alpha} (x,
\tau) = \delta {\cal A}/\delta H_{\alpha} (x, \tau)$. In the Hamiltonian
formalism, this magnetization is measured by the operator $\hat{M}_{\alpha}
(x)$, and we shall focus on the {\em real time\/}, finite $T$ correlation
function
\begin{displaymath}
C_{\alpha\beta}(x,t) = \left\langle e^{i \hat{\cal H} t} \hat{M}_{\alpha} (x)
e^{-i\hat{\cal H} t} \hat{M}_{\beta} (0) \right\rangle - \left\langle\hat{M}_{\alpha}
\right\rangle \left\langle \hat{M}_{\beta} \right\rangle
\end{displaymath}
where $\hat{\cal H}$ is the Hamiltonian corresponding to the action ${\cal
A}$, and the expectation values are with respect to the density matrix
$e^{-\hat{\cal H}/T}/\mbox{Tr} e^{-\hat{\cal H}/T}$. The dimensions of $M$
are inverse length, and because $M$ is a conserved density, it does not
acquire any anomalous dimension ({\em i.e.\/} no prefactors of powers of
$\Lambda$ or $\ln \Lambda$ are required to obtain a finite $\Lambda
\rightarrow
\infty$ limit), and its correlators are simply universal functions of
combinations of $x$, $t$, $c$, $T$, $H$ and $\Delta$ which are consistent
with naive dimensional analysis in lengths and times~\cite{CSY}. For $H \ll
\Delta$ (which we assume throughout), these correlators describe the
crossover between two distinct limiting physical regimes: ({\em i\/}) $T
\ll \Delta$, the `quantum-disordered' regime,  where strong quantum
fluctuations create a paramagnetic ground state, and the excitations
consist of a {\em triplet\/} of particles with energy $(\Delta^2 + c^2 p^2
)^{1/2}$ at momentum $p$, and ({\em ii\/}) $\Delta
\ll T \ll c \Lambda$,
the high $T$ regime of the continuum theory, where quantum fluctuations are
marginally subdominant~\cite{luscher} (by a factor of $1/\ln(T/\Delta)$),
and the excitations are a {\em doublet\/} of spin-waves about a locally
ordered state; however thermal fluctuations of classically interacting
spin-waves lead again to a paramagnetic state. The crossover between these
regimes has been described for the static, uniform, spin susceptibility,
$\chi$, of the $O(N=\infty)$ model~\cite{joli}.

In this paper, we shall obtain the space-time dependent $C(x,t)$ of the
$O(3)$ model in the low $T$ region $T \ll \Delta$. The ratio $H/T$ is
however allowed to be arbitrary. A recent paper~\cite{sagi} computed
$C(0,t)$ by arguing that the triplet of particles could be considered free
at low enough $T$. Actually, such an approach is valid only for $|t|$
shorter than the mean collision time $\sim e^{\Delta/T} / T$ (see below),
and it is essential to include particle collisions at longer $|t|$ to
obtain the crossover to diffusive behavior. Our semiclassical approach does
this, and is valid for all $|t| \gg 1/T$.

\begin{figure}
\epsfxsize=2.0in
\centerline{\epsffile{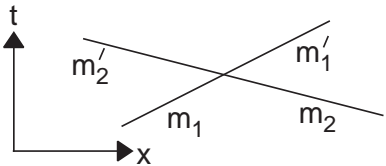}}
\caption{Two particle collision described by the $S$-matrix
(\protect\ref{smatrix}). The momenta before and after the collision are
the same, so the figure also represents the spacetime trajectories of
the particles. }
\label{fig1}
\end{figure}

There are two key observations that allow our exact computation for $T\ll
\Delta$. The first~\cite{apy} is that as there is an excitation gap, the
density of particles $\sim e^{-\Delta/T}$, and their mean spacing is much
larger than their thermal de-Broglie wavelength $\sim c/(\Delta T)^{-1/2}$;
as a result the particles can be treated semiclassically. In particular,
taking the field $H$ pointing along the 3 direction, the density of a
particle with longitudinal spin $m$ ($m = -1,0,1$) is
\begin{displaymath}
\rho_{m} = \int \frac{dp}{2\pi} e^{-(\Delta -m H + c^2 p^2/2 \Delta)/T}
= \sqrt{ \frac{T \Delta}{2\pi c^2} }  e^{-(\Delta - m H)/T},
\end{displaymath}
and therefore the total density $\rho = \rho_{-1} + \rho_0 + \rho_{1}$,
and the magnetization $\langle M_{\alpha} \rangle = (\rho_1 - \rho_{-1})
\delta_{\alpha 3}$.
The second observations is that collisions between these particles are
described by their known two-particle $S$-matrix~\cite{zama}, and only a
simple limit of this $S$-matrix is needed in the low $T$ limit. The r.m.s.
velocity of a thermally excited particle $v_T = c (T/\Delta)^{1/2}$, and
hence its `rapidity' $\sim v_T / c \ll 1$. In this limit, the $S$-matrix
for the process in
Fig~\ref{fig1} is~\cite{zama}
\begin{equation}
 {\cal S}^{m_1 m_2}_{m^{\prime}_1,
m^{\prime}_2} = (-1) \delta_{m_1 m^{\prime}_2} \delta_{m_2 m^{\prime}_1}.
\label{smatrix}
\end{equation}
In other words, the excitations behave like impenetrable
particles which preserve their spin in a collision.
Energy and momentum conservation in $d=1$ require that these
particles simply exchange momenta across a collision (Fig~\ref{fig1}). The
$(-1)$ factor in (\ref{smatrix}) can be interpreted as the phase-shift of
repulsive scattering between slowly moving bosons in $d=1$. Indeed, it
appears that the simple form of (\ref{smatrix}) is due to the slow
motion of the particles, and is not a special feature of relativistic
continuum theory: we conjecture that (\ref{smatrix}) also holds for lattice
Heisenberg spin chains in the limit of vanishing velocities.

\begin{figure}
\epsfxsize=3.0in
\centerline{\epsffile{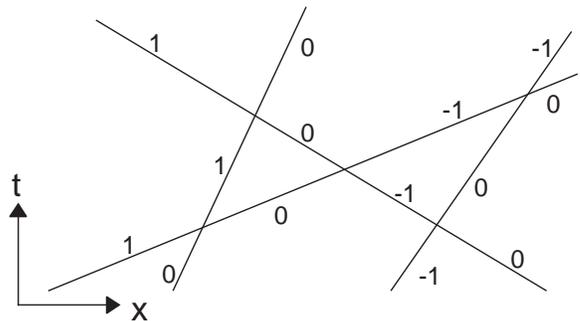}}
\caption{A typical set of particle trajectories contributing to $C(x,t)$.
Each trajectory represents paths moving both forward and backward in time.
The particle co-ordinates are $x_k (t)$, with the labels $k$ chosen so that
$x_k(t) \leq x_l(t)$ for all $t$ and $k < \ell$.
Shown on the trajectories are the values of the particle spins $m_k$ which
are independent of $t$ in the low $T$ limit.}
\label{fig2}
\end{figure}

We now evaluate $C(x,t)$ along the lines of a recent computation for
the Ising model~\cite{apy}. We represent $C(x,t)$ as a `double time'
path integral, with the $e^{-i\hat{\cal H}t}$ factor generating
trajectories that move forward in time, and the $e^{i \hat{\cal H} t}$
producing trajectories that move backward in time. In the classical
limit, stationary phase is achieved when the trajectories are
time-reversed pairs of classical paths (Fig~\ref{fig2}).
Each trajectory has a spin label which obeys (\ref{smatrix}) at each collision;
however as each collision contributes both to the forward and backward
trajectories, the net numerical factor is simply $+1$.
All of this implies~\cite{apy} that the lines in Fig~\ref{fig2} are independently
distributed uniformly in space, and with an inverse slope determined by the velocity $v$
which is distributed according to the
classical Boltzmann probability density
${\cal P}(v) \propto e^{-\Delta v^2/2 c^2 T}$. The spin, $m$, is assigned
randomly at some initial time with
probability $f_m \equiv \rho_{m}/\rho = e^{m H/T}/(1 + 2 \cosh(H/T))$,
but then evolves in time as discussed above (Fig~\ref{fig2}).

We label the particles consecutively from left to right by an integer
$k$ (see the caption of Fig~\ref{fig2});
then their spins $m_k$ are independent of $t$, and we
denote their trajectories $x_k (t)$. The longitudinal correlation $C_{33}$ is given by the
correlators of the classical observable
\begin{equation}
M_3 (x, t) = \sum_k m_k \delta( x - x_k (t))
\label{m3class}
\end{equation}
in the classical ensemble defined above. Now because the spin and
spatial co-ordinates are independently distributed, the correlators
of $m_k$ and $x_k$ factorize. The correlators of the $m_k$ are
easily evaluated:
\begin{equation}
\langle m_k m_{\ell} \rangle  = A_1 + A_2 \delta_{k \ell}
\label{spin}
\end{equation}
where $A_1 \equiv (f_{1} - f_{-1})^2$ and $A_2 \equiv f_{1} + f_{-1} - (f_{1} - f_{-1})^2$
are simple, dimensionless, known functions of $H/T$ only.
Using (\ref{spin}) we have
\begin{eqnarray}
&& C_{33} (x - x', t - t') = A_1 \left( \left\langle \rho(x,t) \rho(x', t')
\right\rangle -
\rho^2\right)
\nonumber \\ &&~~~~~~~~+ A_2 \sum_k \left\langle \delta(x - x_k (t))
\delta(x' - x_k(t'))
\right\rangle
\label{ab}
\end{eqnarray}
where $\rho(x,t) = \sum_k \delta( x- x_k (t))$ is the spacetime dependent
total density, all averages are now with respect to the classical ensemble,
and $\langle \rho(x,t) \rangle = \rho$.
The two-point correlators of $\rho (x,t)$ are also easy to evaluate: if
the spin labels are neglected, the trajectories in Fig~\ref{fig2} are
straight lines, and the density correlators are simply those of a classical ideal gas
of point particles. The second correlator in (\ref{ab}), multiplying $A_2$,
is more difficult: it involves the self correlation a given
particle $k$, which follows a complicated trajectory in the way we
have labeled the particles ({\em e.g.\/} the trajectory of the $-1$ in Fig~\ref{fig2}).
Fortunately, precisely this correlator was considered
three decades ago by Jepsen~\cite{jepsen} and a little later by
others~\cite{lebowitz}; they showed that, at sufficiently long times, each
such particle executes free Brownian motion.
Inserting their results into (\ref{ab}), we obtained the final results
presented below after some straightforward simplifications.

An important property of the results is that they can written in a
`reduced' scaling form~\cite{CSY} determined by the classical dynamics.
From the many independent parameters $c$, $\Delta$, $T$, and $H$,
only a single length ($L_x$)
and a single time ($L_t$) scale controls their spacetime dependence.
These scales can be chosen to be
\begin{equation}
L_x = \frac{1}{\rho} ~~~~~L_t = \frac{1}{\rho} \left( \frac{\Delta}{2
c^2 T} \right)^{1/2}.
\label{defl}
\end{equation}
Notice $L_x \sim c e^{\Delta/T}/\sqrt{\Delta T}$ is the mean spacing between the particles ,
and $L_t \sim e^{\Delta/T}/T$ is a
typical time between particle collisions as  $v_T = L_x / L_t \sqrt{2}$.
Our final result is
\begin{equation}
C_{33} (x,t) = \rho^2 \left[
 A_1 F_1 \left( \frac{|x|}{L_x}
, \frac{|t|}{L_t} \right) + A_2 F_2 \left( \frac{|x|}{L_x} , \frac{|t|}{L_t} \right)
 \right]
\end{equation}
where $\rho^2 F_1$ is the connected density correlator of a classical ideal
gas in $d=1$,
\begin{equation}
F_1 ( \bar{x}, \bar{t} ) = e^{-\bar{x}^2/\bar{t}^2}/\bar{t} \sqrt{\pi},
\end{equation}
and $\rho^2 F_2$ is the correlator of a given labeled
particle~\cite{jepsen,lebowitz},
\begin{eqnarray}
&& F_2 ( \bar{x}, \bar{t} ) =
\Biggl[ \left(2 G_1 (u) G_1 (-u) + F_1( \bar{x},
\bar{t}) \right)  \nonumber \\
&&~~~~~~~~~~~~~~~~\times I_0 \left( 2 \bar{t} \sqrt{G_2 (u) G_2 (-u)} \right)
\nonumber \\
&& ~~~~~~~~~~~~~~~~~+ \frac{G_1^2 (u) G_2 (-u) + G_1^2(-u) G_2 (u)}{\sqrt{G_2(u) G_2 (-u)}}
\nonumber \\
&&~~~~~~~~~~~~\times I_1 \left( 2 \bar{t} \sqrt{G_2 (u) G_2 (-u)} \right) \Biggr]
e^{-(G_2(u) + G_2 (-u)) \bar{t}}
\end{eqnarray}
with $u \equiv \bar{x}/\bar{t}$,
$G_1 (u) = \mbox{erfc} (u)/2$, and
$G_2 (u) = e^{-u^2}/(2 \sqrt{\pi}) - u G_1 (u)$.
These expressions satisfy $
\int_{0}^{\infty} d \bar{x} F_{1,2} (\bar{x}, \bar{t}) = 1/2$,
which ensures the conservation of the total magnetization density with
time. For $|\bar{t}| \ll |\bar{x}| \ll 1$, the function $F_2$ has the ballistic form
$F_2 (\bar{x}, \bar{t} ) \approx F_1 (\bar{x} , \bar{t})$, while for
$|\bar{t}| \gg 1, |x|$ it crosses over to the {\em diffusive \/} form
\begin{equation}
F_2 (\bar{x} , \bar{t}) \approx \frac{e^{-\sqrt{\pi}\bar{x}^2/2 \bar{t}}}{(4 \pi
\bar{t}^2)^{1/4}}~~~~\mbox{for large $\bar{t}$.}
\label{larget}
\end{equation}
In the original dimensionful units, (\ref{defl}) and (\ref{larget})
imply a spin diffusion constant, $D_s$, given exactly by
\begin{equation}
D_s = \frac{c^2 e^{\Delta/T}}{ \Delta (1 + 2 \cosh(H/T))} .
\label{diffres}
\end{equation}

Let us now consider
correlations of the transverse magnetization. It is convenient to work
with the circularly polarized components of the magnetization $M_{\pm} =
M_1 \pm i M_2$. The analog of (\ref{m3class}) is now
$M_{\pm} = \sum_k S_{\pm k} \delta(x - x_k (t))$, where $S_{+k}$
($S_{-k}$) are the spin raising (lowering) operators of particle $k$.
In the double time path integral for $C_{-+} (x,t)$ the spin on some particle
$k$ is raised at time $t=0$ in the forward trajectory;
at time $t$ the lowering operator must act
on the {\em same\/} particle or otherwise the trace over the classical
trajectories vanishes. Notice also that there is no raising or lowering
of spins in the backward trajectory. As a result, the path integral picks
up a factor of $e^{i H t}$ from the additional Berry
phase accumulated during the time the spin is raised during the forward
trajectory, which is not compensated by the backward trajectory.
This phase is multiplied by the self-correlation of the particle whose
spin was raised, a quantity we have obtained above.
Similar considerations apply to $C_{+-}$ and the final
results are
\begin{equation}
C_{\mp\pm}(x,t) = 2 \rho^2 e^{\pm i H t} A_{\mp}
F_2 \left( \frac{|x|}{L_x}
, \frac{|t|}{L_t} \right)
\end{equation}
where $A_{\mp} \equiv f_0 + f_{\mp 1}$.

\begin{figure}
\epsfxsize=2.5in
\centerline{\epsffile{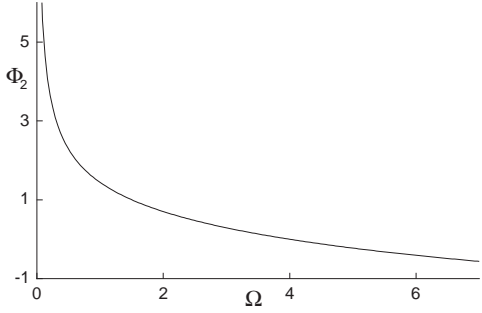}}
\caption{The crossover function $\Phi_2 (\Omega)$ appearing in the local
dynamic structure factor.}
\label{fig3}
\end{figure}

Next, we compute
the local dynamic structure factor
$ S_{\alpha\beta} (\omega ) = \int_{-\infty}^{\infty} dt e^{-i\omega t} C_{\alpha\beta} (0,
t)$. A subtlety arises in computing this Fourier transform. Notice
that at short $|t| \ll L_t$, we have the ballistic behavior $C(0,t) \sim 1/|t|$, and so the
$t$ integral is logarithmically divergent. Our semiclassical results are valid
only for $|t| \gg 1/T$, and so we should cut-off the integral at small $t$,
leading to a contribution $\sim \ln (b T/\omega)$ where $b$ is a numerical factor of order
unity. In fact, it is possible to determine $b$ precisely: at these short times the earlier
free quantum particle approach~\cite{sagi} is valid, and we determine $b$ by matching the
logarithm to their results. In physical terms, the short time cut-off is provided by the
wavelike nature of the individual particles, at a scale where collisions are unimportant.
Our final results for $S( \omega)$ are
\begin{eqnarray}
S_{33} ( \omega ) &&=
\frac{\rho}{c} \sqrt{\frac{2  \Delta}{\pi T}} \left[
A_1 \left\{ \ln (T L_t) + \Phi_1 ( \sqrt{\pi} |\omega| L_t) \right\} \right. \nonumber \\
&&~~~~~~~~~\left.+ A_2 \left\{ \ln (T L_t) + \Phi_2 ( \sqrt{\pi} |\omega| L_t ) \right\} \right]
\nonumber
\\ S_{\mp \pm} ( \omega ) &&=
\frac{2\rho A_{\mp}}{c} \sqrt{\frac{2  \Delta}{\pi T}} \left\{ \ln (T L_t) + \Phi_2 (
\sqrt{\pi} |\omega\mp H| L_t) \right\} \nonumber
\end{eqnarray}
The $\ln (T L_t)$ terms logarithmically violate the purely classical,
reduced scaling forms~\cite{CSY},
and were fixed by matching to the short-time quantum calculation~\cite{sagi}.
The scaling functions $\Phi_{1,2} (\Omega )$ were determined to be
\begin{eqnarray}
\Phi_1 ( \Omega ) &&= \ln \left( \frac{4 \sqrt{\pi} e^{-\gamma}}{\Omega} \right)
\nonumber \\
\Phi_2 (\Omega ) &&= \Phi_1 ( \Omega ) + \frac{\pi [ ( \sqrt{4 + \Omega^2} + 2)^{1/2} -
\sqrt{\Omega} ]^2}{ 4 \sqrt{\Omega} ( \sqrt{4 + \Omega^2} + 2)^{1/2} } \nonumber \\
&&~~~~ - \ln \frac{( 1 + \Omega^2
/\Psi^2(\Omega))^{1/2}  (1 + \Psi(\Omega)) }{2\Omega}
\label{psis}
\end{eqnarray}
where $\gamma = 0.57721\ldots$ is Euler's constant, and $\Psi ( \Omega ) =
( \Omega \sqrt{1 + \Omega^2 / 4} - \Omega^2 / 2 )^{1/2}$.
We show a plot of the scaling function $\Phi_2 ( \Omega)$ in Fig~\ref{fig3}: it clearly shows
the expected crossover from the large frequency ballistic behavior $\Phi_2 ( \Omega \rightarrow
\infty )  = \ln ( 1/\Omega)$, to the small frequency diffusive form
$\Phi_2 ( \Omega \rightarrow 0 ) = \pi /(2 \sqrt{ \Omega})$.

The longitudinal relaxation rate of nuclei coupled to the electronic spins
is $1/T_1 = (\Gamma /2) S_{+-} (\omega_N ) $, where $\Gamma$ is determined
by the electron-nucleus hyperfine coupling, and $\omega_N$ is a nuclear
frequency which can safely be set to zero. It is useful to explicitly note
the $H \ll L_t^{-1}$ limit of $1/T_1$, where from (\ref{psis}), we
have
\begin{displaymath}
\frac{1}{T_1} = \frac{\Gamma T \chi}{\sqrt{2 D_s H}} =
\frac{\Gamma \Delta e^{-3 \Delta / 2 T}}{c^2} \sqrt{\frac{3 T}{\pi
H}} ~~;~~\chi = \frac{e^{-\Delta / T}}{c} \sqrt{ \frac{2 \Delta}{ \pi T}},
\end{displaymath}
where $\chi = \lim_{H \rightarrow 0} (\rho_1 - \rho_{-1})/H$ was known
earlier~\cite{tsvelik,sagi}. For experimental comparisons, an important
property of the above, pointed out to us by M.~Takigawa, is that the low
$T$ activation gaps for $1/T_1$ ($\Delta_{1/T_1}$) and $\chi$
($\Delta_{\chi}$) satisfy $\Delta_{1/T_1}/\Delta_{\chi} = 3/2$.

A quantitative comparison of our results with experiments requires a
detailed study of the $H$ and $T$ dependencies of $1/T_1$, along with
consideration of effects due to spin-anisotropies and inter-chain couplings
which can become important at low $T$ and $H$. Such an analysis will be
presented elsewhere; here, we simply note some trends which appear to
receive a natural explanation from our theory. Values for the activation
gaps $\Delta_{1/T_1}$ and $\Delta_{\chi}$ have been quoted for a number of
experimental systems~\cite{taki,shimizu,azuma}, and it has consistently
been found that $\Delta_{1/T_1}$ is larger than $\Delta_{\chi}$. In  the
spin $S=1$ chain compound $Ag V P_2 S_6$ Takigawa {\em et.
al.}~\cite{taki} estimated $\Delta_{1/T_1} /
\Delta_{\chi} = 1.3$; for the spin $S=1$ chain compound $Y_2 Ba Ni O_5$,
Shimizu {\em et. al.\/}~\cite{shimizu} measured
$\Delta_{1/T_1}/
\Delta_{\chi} = 1.53 \pm 0.08$; finally, in the two-leg $S=1/2$ ladder
compound $Sr Cu_2 O_3$, Azuma {\em et. al.\/}~\cite{azuma} found
$\Delta_{1/T_1}/\Delta_{\chi} = 1.6$.

Takigawa {\em et. al.\/}~\cite{taki} also observed the diffusive
$1/\sqrt{H}$ dependence of $1/T_1$, from which the  value of $D_s$ was
estimated: $D_s /a^2
\approx 5.5\times 10^{14}$ $\mbox{sec}^{-1}$ at $T=220 K$, where $a$ is the
lattice spacing. From measurements~\cite{taki2} of $\chi$ we may obtain
$\Delta = 320 K$, and $c/a = 3.32 \Delta$, which when inserted into
(\ref{diffres}) give $D_s/a^2 = 6.6 \times 10^{14}$ $\mbox{sec}^{-1}$.
However, it should be noted that numerical analysis~\cite{soren2} on the
nearest-neighbor $S=1$ antiferromagnet gives $c/a = 6.06 \Delta$, but using
this value of $c$ would also lead to a discrepancy in the theoretical
prediction for $\chi$.

Finally, we note that similar methods~\cite{apy} can be used to obtain
dynamic, $T > 0$, correlators of the $n_{\alpha}$ field: this will be
described elsewhere.

We are indebted to M. Takigawa for surveying and interpreting the
experimental situation for us. We thank Satya N.~Majumdar and T.~Senthil
for stimulating discussions which provoked our interest in this problem,
and I.~Affleck for helpful remarks. This research was supported by NSF
Grant No DMR 96--23181.

\end{document}